\documentclass[12pt]{article}
\usepackage{graphicx}
\usepackage{amssymb}
\usepackage{amsmath}
\usepackage{subfigure}
\input{colordvi.tex}

\usepackage{color}
\usepackage[colorlinks=true,linkcolor=blue,citecolor=blue]{hyperref}

\begin{document}
\baselineskip 0.65cm

\begin{titlepage}

\begin{flushright}
ICRR-Report-674-2013-23\\
IPMU14-0066 
\end{flushright}

\vskip 1.35cm

\begin{center}

{\large 
{\bf Compensation for large tensor modes with iso-curvature perturbations
 in CMB anisotropies} 
}

\vskip 1.2cm

Masahiro Kawasaki$^{a,b}$
and
Shuichiro Yokoyama$^a$ \\

\vskip 0.4cm

{ \it$^a$Institute for Cosmic Ray Research,
University of Tokyo, Kashiwa 277-8582, Japan}\\
{\it$^b$ Kavli Institute for the Physics and Mathematics of the
Universe (Kavli IPMU), TODIAS,  the University of Tokyo, 5-1-5
Kashiwanoha, Kashiwa, 277-8583, Japan}\\

\date{\today}

\vspace*{1.5cm}
\begin{abstract} 
Recently, BICEP2 has reported the large tensor-to-scalar ratio $r=0.2^{+0.07}_{-0.05}$
from the observation of the cosmic microwave background (CMB) B-mode at degree-scales.
Since tensor modes induce not only CMB B-mode but also the temperature fluctuations on large scales,
to realize the consistent temperature fluctuations with the Planck result
we should consider suppression of scalar perturbations on corresponding large scales.
To realize such a suppression, we consider anti-correlated iso-curvature perturbations
which could be realized in the simple curvaton model.   
\end{abstract}

\end{center}
\end{titlepage}

\section{Introduction}
\label{sec:intro}

Recently, BICEP2 collaboration has reported the detection of the cosmic microwave background (CMB)
B-mode polarization at the degree scales at $7.0\sigma$~\cite{Ade:2014xna,Ade:2014gua}. The detected degree scale B-mode polarization
is consistent with the primordial gravitational waves origin and the corresponding value of the tensor-to-scalar ratio, which represents the amplitude of the primordial gravitational waves,  
is $r=0.2^{+0.07}_{-0.05}$. 
The amplitude of the primordial gravitational wave is a direct probe of the energy scale of the inflation,
which is given by
\begin{eqnarray}
H_{\rm inf} \simeq 1.22 \times 10^{14}~{\rm GeV} \left( {r \over 0.2}\right)^{1/2}.
\end{eqnarray}
Based on this result, there have appeared a number of interesting papers~\cite{Zhao:2014rna,Nakayama:2014koa,Higaki:2014ooa,Marsh:2014qoa,Harigaya:2014sua,Anchordoqui:2014uua,Ma:2014vua,Czerny:2014wua,Byrnes:2014xua,Collins:2014yua,Visinelli:2014twa,Contaldi:2014zua,Harigaya:2014qza,Kehagias:2014wza,Giusarma:2014zza,Lizarraga:2014eaa,Brandenberger:2014faa,Choudhury:2014kma,Ashoorioon:2014nta,Biswas:2014kva}.

However, there is a tension between BICEP2 and Planck results.
The latter gives the upper bound on the
tensor-to-scalar ratio as $r < 0.11$ obtained from the combined analysis of the CMB temperature fluctuations
and other data sets~\cite{Ade:2013zuv,Ade:2013uln}.
The key point to resolve this discrepancy is a suppression of the scalar perturbations on large scales enough to
compensate for the contribution from the large tensor modes.
As an example, running spectral index is discussed in
BICEP2 paper and it is necessary to consider relatively large (in terms of absolute value) negative
running such as $d n_s / d\ln k \sim - 0.02$ to resolve the tension, which is much larger than
the expected value in the standard slow-roll inflation model.
In Ref. \cite{Contaldi:2014zua}, the authors consider the anti-correlation between the scalar and tensor perturbations instead of the running spectral index. 

In this paper, we alternatively consider the anti-correlated iso-curvature perturbation
to compensate for the contribution of the large tensor modes~\cite{Kawasaki:2007mb,Valiviita:2012ub,Savelainen:2013iwa}.
Such iso-curvature perturbations could be realized in the simple curvaton model
which has both cold dark matter (CDM) and baryon isocurvature perturbations.
This paper is organized as follows.
In the next section, we show our basic idea of compensation for the tensor contribution with the iso-curvature perturbations.
In section \ref{curvaton}, 
in the context of the curvaton scenario
we present a concrete model where the required 
iso-curvature perturbation could be realized,
and we devote the final section to summary.

\section{Compensation caused by iso-curvature perturbations}
\label{sec:setup}

In this section let us describe the basic idea.
We focus on the compensation for the large tensor modes 
due to the iso-curvature perturbations.
Owing to the damping behavior of the tensor and iso-curvature perturbations,
both perturbations give similar contributions to large scale temperature fluctuations.  
On large scales, the temperature anisotropy induced from scalar perturbation is mainly sourced from the Sachs-Wolfe effect as
\begin{eqnarray}
\left( \Delta T \over T \right)_{\rm SW} = -{1 \over 5} \zeta - {2 \over 5} {\mathcal S}_{\rm m},
\end{eqnarray}
where $\zeta$ is the adiabatic curvature perturbation on uniform energy density slice during the radiation-dominated era,
and ${\mathcal S}_{\rm m}$ is the total matter iso-curvature perturbations.
The matter component consists of the CDM and baryons and hence we have
\begin{eqnarray}
{\mathcal S}_{\rm m} = {\Omega_{\rm CDM} \over \Omega_{\rm m} }
{\mathcal S}_{\rm CDM} + {\Omega_{\rm b} \over \Omega_{\rm m}} {\mathcal S}_{\rm b},
\end{eqnarray}
where ${\mathcal S}_{\rm CDM}$ and ${\mathcal S}_{\rm b}$ are respectively CDM and baryon iso-curvature
perturbations, and $\Omega_i$ ($i=$ CDM, b, and m)  with $\Omega_{\rm m} = \Omega_{\rm CDM} + \Omega_{\rm b}$ is a density parameter of each component.
As we have mentioned, the tensor perturbations also contribute the CMB temperature fluctuations.
On large scales, the tensor-contribution can be approximately written as~\cite{Contaldi:2014zua}
\begin{eqnarray}
\left( \Delta T \over T \right)_{\rm tens} \simeq {1 \over 2} h_{ij} n^i n^j,  
\end{eqnarray}
where $n$ is a unit vector.
Following Ref. \cite{Contaldi:2014zua},
including all contributions to the temperature anisotropies
we obtain 
\begin{eqnarray}
\langle \left({\Delta T \over T}\right)^2\rangle
&\propto& 
{{\mathcal P}_\zeta} \left( 1 + 4 {{\mathcal P}_{{\mathcal S}_{\rm m}} \over {\mathcal P}_\zeta} 
+ 4 {{\mathcal P}_{\zeta {\mathcal S}_{\rm m} } \over {\mathcal P}_\zeta} 
+ {5 \over 6} {{\mathcal P}_T \over {\mathcal P}_\zeta}\right) \cr\cr
&=& {\mathcal P}_\zeta \left( 1 + 4 B_{\rm m}^2 + 4 B_{\rm m} \cos \theta_{\rm m} +
{1 \over 6} \left( {r \over 0.2} \right)
\right),
\label{eq:ratio}
\end{eqnarray}
where
${\mathcal P}_i$ ($i = \zeta, {\mathcal S}_{\rm m}, T$) is the power spectrum of each component, and 
${\mathcal P}_{\zeta{\mathcal S}_{\rm m}}$ is a cross-power spectrum of the adiabatic and iso-curvature
perturbations.
Here, we also introduce parameters related with the iso-curvature perturbations as
\begin{eqnarray}
B_{\rm m} \equiv \sqrt{{{\mathcal P}_{{\mathcal S}_{\rm m}} \over {\mathcal P}_\zeta}}, ~
\cos \theta_{\rm m}  \equiv  {{\mathcal P}_{\zeta {\mathcal S}_{\rm m}} \over \sqrt{{\mathcal P}_\zeta {\mathcal P}_{{\mathcal S}_{\rm m}}}} .
\end{eqnarray}
From Eq.~(\ref{eq:ratio}), to compensate for the tensor contribution with the iso-curvature perturbation,
the following relation is required:
\begin{eqnarray}
4 B_{\rm m}^2 +4 B_{\rm m} \cos \theta_{\rm m} +
{1 \over 6} \left( {r \over 0.2} \right) = 0.
\label{eq:relation}
\end{eqnarray} 
To realize this requirement, we at least need anti-correlated iso-curvature perturbations, that is,
$\cos \theta_{\rm m} < 0 $, and with $\cos \theta_{\rm m} = -1$ we obtain
\begin{eqnarray}
B_{\rm m} = {1 \over 2} \left( 1 \pm \sqrt{1-{1 \over 6} \left( {r \over 0.2}\right)} \right) > 0.
\end{eqnarray}
If we chose the BICEP2 result, $r=0.2$, we have $B_{\rm m} \simeq 0.04$ or $0.96$.
In Ref.~\cite{Kawasaki:2007mb}\footnote{We would like to note that there are related papers about the observational constraint on the iso-curvature perturbations
 with including the contribution from the tensor perturbations by using WMAP 7 \cite{Valiviita:2012ub} and WMAP 9 data \cite{Savelainen:2013iwa}.
 Although these papers used different parameterization from that used in Ref. \cite{Kawasaki:2007mb},
 obtained results are consistent.}, by using the WMAP3 and other data sets
the constraints on the value of $B_{\rm m}$ and $r$ are respectively given by
$ B_{\rm m} \cos \theta_{\rm m} \geq  - 0.13$ and $r \leq 0.49$. 
Hence, our naive estimation seems to be agreement with these constraints 
for $B_{\rm m} = 0.04$.

\section{How to realize required anti-correlated iso-curvature perturbations?}
\label{curvaton}

Let us consider the concrete example which can realize the required anti-correlated iso-curvature
perturbations as we have shown in the previous section.
Here, we consider the curvaton scenario~\cite{Enqvist:2001zp,Lyth:2001nq,Moroi:2001ct}
as the mechanism of generating the iso-curvature perturbations \cite{Moroi:2002rd,Lyth:2002my,Lyth:2003ip,Ferrer:2004nv,Beltran:2008aa,Moroi:2008nn,Lemoine:2009is}.

In the curvaton scenario, the adiabatic curvature perturbations could be induced not only from the inflation
fluctuations but also fluctuations of a light scalar field other than inflaton called as curvaton.
The curvaton field which is subdominant component of the Universe has acquired the fluctuations during inflation, and then starts to oscillate when the Hubble
parameter becomes equal to the mass of the field. Such oscillating scalar field behaves as non-relativistic matter. Hence, during radiation dominated era, the energy density of the curvaton becomes larger
compared with the total energy density of the Universe and
simultaneously the adiabatic fluctuations also evolves due to the iso-curvature perturbations
sourced from the curvaton fluctuations.
Finally, after the curvaton decays the adiabatic curvature perturbations stay constant in time.
The total adiabatic curvature perturbations after the curvaton decay is given by
\begin{eqnarray}
\zeta = \zeta_{\rm inf} + {f_{\rm dec} \over 3} {\mathcal S}_{\sigma},
\label{eq:curv_pert}
\end{eqnarray}
where 
$f_{\rm dec} (=3\rho_\sigma / (3\rho_\sigma + 4\rho_r)|_{\rm decay})$ is a parameter related with the ratio between the curvaton and the radiation
energy density at the curvaton decay,
$\zeta_{\rm inf}$ is the curvature perturbations induced from the inflaton fluctuations
and ${\mathcal S}_{\sigma}$ is the curvaton iso-curvature perturbations which is given by
${\mathcal S}_{\sigma} = 3 (\zeta_\sigma - \zeta_{\rm inf})$ with $\zeta_{\sigma}$ being
the curvature perturbations on the uniform energy density slicing of the curvaton.

As for the residual iso-curvature perturbations, ${\mathcal S}_{\rm CDM}$ or ${\mathcal S}_{\rm b}$,
it depends on the generation process of the CDM and baryons.
In the case where the CDM/baryons have been produced from the inflaton decay,
the curvature perturbation on the uniform CDM/baryons energy density slice, $\zeta_{\rm CDM/b}$,
should be equal to $\zeta_{\rm inf}$ and then the residual iso-curvature perturbations 
defined as ${\mathcal S}_{\rm CDM/b} \equiv 3 (\zeta_{\rm CDM/b} - \zeta)$ are given by
\begin{eqnarray}
{\mathcal S}_{\rm CDM/b} = - f_{\rm dec} {\mathcal S}_{\sigma} .
\end{eqnarray}
From this expression, we find that this case has the possibility of generating required anti-correlated
iso-curvature perturbations. 
On the other hand, in the case where the CDM/baryons have been produced from the curvaton
decay, we have $\zeta_{\rm CDM/b} = \zeta_{\sigma}$ and then the residual iso-curvature perturbations
are given by
\begin{eqnarray}
{\mathcal S}_{\rm CDM/b} = \left(1 - f_{\rm dec} \right) {\mathcal S}_{\sigma} .
\end{eqnarray}
In this case, it is difficult to realize the anti-correlated iso-curvature perturbations
because of the fact that $f_{\rm dec} < 1$.

Let us focus on the former case and assume that both of CDM and baryons have been produced from
the inflaton decay. 
In this case, the parameters introduced in the previous section are respectively
written as
\begin{eqnarray}
B_{\rm m} = \sqrt{ f_{\rm dec}^2 R \over \left( 1 + {f_{\rm dec}^2 \over 9} R\right) },
~\cos \theta_{\rm m} = -  {f_{\rm dec}^2  R \over 3 \sqrt{ \left( 1 + {f_{\rm dec}^2 \over 9} R\right) f_{\rm dec}^2 R}},
\end{eqnarray}
where $R$ represents the ratio of the amplitude of ${\mathcal S}_{\sigma}$ to that of $\zeta_{\rm inf}$
defined as $R \equiv  {\mathcal P}_{{\mathcal S}_\sigma} / {\mathcal P}_{\zeta_{\rm inf}} $.
Substituting the above expressions into the left hand side of Eq. (\ref{eq:relation}), we have
\begin{eqnarray}
4 B_{\rm m}^2 +4 B_{\rm m} \cos \theta_{\rm m} +
{1 \over 6} \left( {r \over 0.2} \right) 
= {8 \over 3} {f_{\rm dec}^2 R \over 1 + {f_{\rm dec}^2 \over 9} R} + {1 \over 6} 
\left( {r \over 0.2}\right).
\end{eqnarray}
Since all parameters are positive definite,
this case can not realize the required iso-curvature perturbations.
As we have mentioned, in another case where both of CDM and baryons have been produced from the curvaton decay
the expected residual iso-curvature perturbations cannot be the required one.

\begin{figure}[htbp]
\begin{center}
\includegraphics[width=110mm]{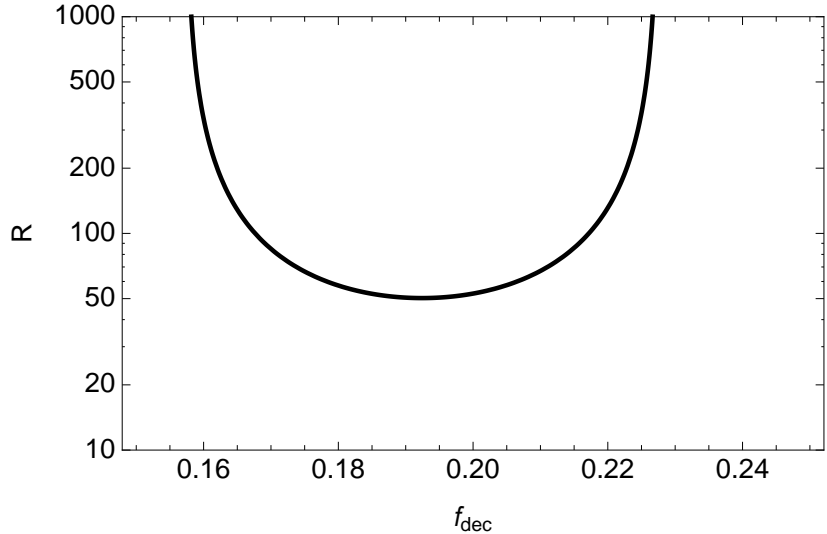}
\end{center}
\caption{
The solution for Eq. (\ref{eq:final}) in $f_{\rm dec}$-$R$ plane.}
\label{fig:fig}
\end{figure}

Next, let us consider the mixed scenario where the CDM and baryons are sourced from the inflaton  and curvaton decay, respectively.
In such case, we have
\begin{eqnarray}
{\mathcal S}_{\rm CDM} = -f_{\rm dec} {\mathcal S}_\sigma,~
{\mathcal S}_{\rm b} = (1 - f_{\rm dec}) {\mathcal S}_\sigma,
\end{eqnarray} 
and the total matter iso-curvature perturbations are given by
\begin{eqnarray}
{\mathcal S}_{\rm m} = \left[ -f_{\rm dec} {\Omega_{\rm CDM} \over \Omega_{\rm m}} 
+ (1 - f_{\rm dec}) \left(1 - {\Omega_{\rm CDM} \over \Omega_{\rm m}} \right) \right] {\mathcal S}_\sigma.
\label{eq:compensated}
\end{eqnarray}
Taking the parameter $f_{\rm dec}$ to be equal to $1 - \Omega_{\rm CDM}/\Omega_{\rm m}$,
then the total matter iso-curvature perturbations is completely cancelled
even as each component has the iso-curvature perturbations, so-called compensated iso-curvature perturbations
\cite{Holder:2009gd,Gordon:2009wx,Kawasaki:2011ze,Grin:2011tf,Grin:2013uya}.
On the other hand, to realize the compensation for the large tensor mode with the iso-curvature perturbation
we do not need completely compensated iso-curvature perturbations
but the partially compensated ones.
By using the above expression Eq. (\ref{eq:compensated}),
we have
\begin{eqnarray}
B_{\rm m} = \sqrt{ {\left( 1 - {\Omega_{\rm CDM} \over \Omega_{\rm m}} - f_{\rm dec} \right)^2 R 
\over 
\left( 1 + {f_{\rm dec}^2 \over 9} R \right)
}
}, ~
\cos \theta_{\rm m} = {f_{\rm dec} \left( 1 - {\Omega_{\rm CDM} \over \Omega_{\rm m}} - f_{\rm dec} \right)
R 
\over 3 \sqrt{\left( 1 + {f_{\rm dec}^2 \over 9} R \right) \left( 1 - {\Omega_{\rm CDM} \over \Omega_{\rm m}} - f_{\rm dec} \right)^2 R }} .
\end{eqnarray}
Substituting the above expression into the left hand side of Eq. (\ref{eq:relation}), we have
\begin{eqnarray}
4{ \left( 1 - {\Omega_{\rm CDM} \over \Omega_{\rm m}} - f_{\rm dec} \right)
R \over 1 + {f_{\rm dec}^2 \over 9} R}
\left( 1 - {\Omega_{\rm CDM} \over \Omega_{\rm m}} - {2 \over 3} f_{\rm dec} \right)
+ {1 \over 6} \left( {r \over 0.2}\right)  = 0.
\label{eq:final}
\end{eqnarray}
As shown in Fig. \ref{fig:fig}, we can easily find that the above equation has solutions for large $R$ case.
%
%
In the limiting case where the curvaton contribution dominates over the adiabatic perturbations, that is,
$f_{\rm dec}^2 R / 9 \gg 1$,
for the Planck best-fit values of $\Omega_{\rm CDM}$ and $\Omega_{\rm m}$ and the BICEP2 value
of $r$, that is, $\Omega_{\rm CDM} / \Omega_{\rm m} = 0.845$ and $r=0.2$,
we find that $f_{\rm dec} \simeq 0.16$ and $0.23$ are solutions for Eq.~(\ref{eq:final}).
Hence, in such case we can realize the required iso-curvature perturbations to compensate for the tensor contribution to the temperature anisotropies, which makes the BICEP2 result consistent with the Planck result.
An intuitive understanding of the need of the compensated isocurvature perturbations is as follows.
In the simpler scenarios where all of the CDM and baryons come from either only the curvaton, or only the inflaton,
the contributions from the curvaton fluctuations to the adiabatic perturbations and the matter iso-curvature perturbations 
are almost same, which are characterized by a single parameter $f_{\rm dec}$.
However, to realize the required iso-curvature perturbations in the light of BICEP2 and Planck,
it should be necessary to consider the larger contribution to the adiabatic perturbations than that to the matter iso-curvature perturbations.
By considering the partially compensated situation, we can introduce a new parameter related with the ratio between the energy density of the CDM and that of the baryon,
and the contribution to the iso-curvature parameter depends on not only $f_{\rm dec}$ but also the new parameter $\Omega_{\rm CDM} / \Omega_{\rm m}$.
Hence, to choose appropriate values of these parameters 
we can realize the case with the larger contribution to the adiabatic perturbations than that to the matter iso-curvature perturbations.

As is well-known, although the parameter $f_{\rm dec}$ is an important parameter to describe the primordial non-Gaussianity and it is strongly constrained as $f_{\rm dec} > 0.15$~\cite{Ade:2013ydc},
this constraint is consistent with the required values $f_{\rm dec} \simeq 0.16$ and $0.23$. 

We would like to note that
the BICEP2 result basically indicates that even if we consider the curvaton scenario
we hardly neglect the inflaton contribution due to the large tensor-to-scalar ratio~\cite{Byrnes:2014xua}.
As shown in Fig. \ref{fig:fig}, for minimum $R$,
 we can also have a solution at $R \simeq 50$ 
and $f_{\rm dec} \simeq 0.19$.
For these parameters, the ratio $R_{\rm curv}$ of the power spectra of 
the curvature perturbations from the inflaton 
to the curvaton is given by $R_{\rm curv}=f_{\rm dec}^2 R / 9 \simeq 0.2$ 
[see, Eq.~(\ref{eq:curv_pert})] 
and hence we can realize the required iso-curvature fluctuations in case 
where the the curvaton contributions are only $20\%$ of the inflaton contributions.\footnote{%
In case $R_{\rm curve} \gtrsim 1$ the slow-roll parameter 
$\epsilon = \dot{H}_{\rm inf}/H_{\rm inf}^2$ ($H_{\rm inf}$:Hubble during inflation)
should be large to suppress the inflaton contribution, which leads to red-tilted 
spectral index of the curvaton perturbations.}

We find that the required iso-curvature perturbations can be also realized for the opposite case,
that is, the case where CDM and baryons are sourced from the curvaton and inflaton decay, respectively.
In such case, the solution can be obtained for relatively large $f_{\rm dec}$ and small $R$.
For example,  we have a solution for $f_{\rm dec} \simeq 0.95 $ and $R \simeq 2$,
and this means $R_{\rm curve} \simeq 0.2$.
Hence, in this case we can also realize the required situation by considering  only $20\%$ curvaton contributions.

\section{Summary}
\label{summary}

Recent BICEP2 report about the detection of the degree-scale CMB B-mode polarization
has revealed the energy scale of the inflation and opened a window into the deep understanding
of the physics of the early Universe.

However there still exist a tension between the BICEP2 result $r=0.2^{+0.07}_{-0.05}$
and Planck result $r< 0.1$ from the observations of the CMB temperature fluctuations.
In order to resolve this discrepancy, we need to consider the suppression of the large scale scalar
perturbations.
In this paper, instead of considering the large negative running discussed in the BICEP2 paper,
we consider the anti-correlated iso-curvature perturbations.
We present a concrete example to realize the required anti-correlated iso-curvature perturbations
in the context of the curvaton scenario.
We just show the possibility by using the rough estimate and hence
it should be interesting to perform more detailed analysis by using e.g., Markov Chain Monte Carlo methods.
We leave such detailed analysis as a future issue.

\section*{Acknowledgments}

We thank Taku Hayakawa for pointing out our typos.
The work is supported by Grant-in-Aid for Scientific Research 25400248(MK), 24-2775 (SY)
and 21111006 (MK) from the Ministry of Education, Culture, Sports, Science and Technology in Japan,
and also by World Premier International Research Center Initiative (WPI Initiative), MEXT, Japan.

{}



\begin{thebibliography}{9}


\bibitem{Ade:2014xna} 
  P.~A.~R.~Ade {\it et al.}  [BICEP2 Collaboration],
  arXiv:1403.3985 [astro-ph.CO].

\bibitem{Ade:2014gua} 
  P.~A.~RAde {\it et al.}  [ BICEP2 Collaboration],
  arXiv:1403.4302 [astro-ph.CO].
  
  
\bibitem{Zhao:2014rna} 
  W.~Zhao, C.~Cheng and Q.~-G.~Huang,
  arXiv:1403.3919 [astro-ph.CO].
  
\bibitem{Nakayama:2014koa} 
  K.~Nakayama and F.~Takahashi,
  arXiv:1403.4132 [hep-ph].
  
\bibitem{Higaki:2014ooa} 
  T.~Higaki, K.~S.~Jeong and F.~Takahashi,
  arXiv:1403.4186 [hep-ph].
  
\bibitem{Marsh:2014qoa} 
  D.~J.~E.~Marsh, D.~Grin, R.~Hlozek and P.~G.~Ferreira,
  arXiv:1403.4216 [astro-ph.CO].
  
\bibitem{Harigaya:2014sua} 
  K.~Harigaya, M.~Ibe, K.~Schmitz and T.~T.~Yanagida,
  arXiv:1403.4536 [hep-ph].
  
\bibitem{Anchordoqui:2014uua} 
  L.~A.~Anchordoqui, V.~Barger, H.~Goldberg, X.~Huang and D.~Marfatia,
  arXiv:1403.4578 [hep-ph].
  
\bibitem{Ma:2014vua} 
  Y.~-Z.~Ma and Y.~Wang,
  arXiv:1403.4585 [astro-ph.CO].
  
\bibitem{Czerny:2014wua} 
  M.~Czerny, T.~Kobayashi and F.~Takahashi,
  arXiv:1403.4589 [astro-ph.CO].

\bibitem{Byrnes:2014xua} 
  C.~T.~Byrnes, M.~Cortes and A.~R.~Liddle,
  arXiv:1403.4591 [astro-ph.CO].
  
\bibitem{Collins:2014yua} 
  H.~Collins, R.~Holman and T.~Vardanyan,
  arXiv:1403.4592 [hep-th].

\bibitem{Visinelli:2014twa} 
  L.~Visinelli and P.~Gondolo,
  arXiv:1403.4594 [hep-ph].

\bibitem{Contaldi:2014zua} 
  C.~R.~Contaldi, M.~Peloso and L.~Sorbo,
  arXiv:1403.4596 [astro-ph.CO].
 
\bibitem{Harigaya:2014qza} 
  K.~Harigaya and T.~T.~Yanagida,
  arXiv:1403.4729 [hep-ph].
  
\bibitem{Kehagias:2014wza} 
  A.~Kehagias and A.~Riotto,
  arXiv:1403.4811 [astro-ph.CO].
  
\bibitem{Giusarma:2014zza} 
  E.~Giusarma, E.~Di Valentino, M.~Lattanzi, A.~Melchiorri and O.~Mena,
  arXiv:1403.4852 [astro-ph.CO].
  
  
\bibitem{Lizarraga:2014eaa} 
  J.~Lizarraga, J.~Urrestilla, D.~Daverio, M.~Hindmarsh, M.~Kunz and A.~R.~Liddle,
  arXiv:1403.4924 [astro-ph.CO].
  
  
\bibitem{Brandenberger:2014faa} 
  R.~H.~Brandenberger, A.~Nayeri and S.~P.~Patil,
  arXiv:1403.4927 [astro-ph.CO].
 
\bibitem{Choudhury:2014kma} 
  S.~Choudhury and A.~Mazumdar,
  arXiv:1403.5549 [hep-th].
 
\bibitem{Ashoorioon:2014nta} 
  A.~Ashoorioon, K.~Dimopoulos, M.~M.~Sheikh-Jabbari and G.~Shiu,
  arXiv:1403.6099 [hep-th].
 
\bibitem{Biswas:2014kva} 
  T.~Biswas, T.~Koivisto and A.~Mazumdar,
  arXiv:1403.7163 [hep-th].
 
 
\bibitem{Ade:2013zuv} 
  P.~A.~R.~Ade {\it et al.}  [Planck Collaboration],
  arXiv:1303.5076 [astro-ph.CO].
  
\bibitem{Ade:2013uln} 
  P.~A.~R.~Ade {\it et al.}  [Planck Collaboration],
  arXiv:1303.5082 [astro-ph.CO].
 
  
\bibitem{Kawasaki:2007mb} 
  M.~Kawasaki and T.~Sekiguchi,
  Prog.\ Theor.\ Phys.\  {\bf 120}, 995 (2008)
  [arXiv:0705.2853 [astro-ph]].
  

\bibitem{Valiviita:2012ub} 
  J.~Valiviita, M.~Savelainen, M.~Talvitie, H.~Kurki-Suonio and S.~Rusak,
  Astrophys.\ J.\  {\bf 753}, 151 (2012)
  [arXiv:1202.2852 [astro-ph.CO]].

  
  
\bibitem{Savelainen:2013iwa} 
  M.~Savelainen, J.~Valiviita, P.~Walia, S.~Rusak and H.~Kurki-Suonio,
  Phys.\ Rev.\ D {\bf 88}, 063010 (2013)
  [arXiv:1307.4398 [astro-ph.CO]].
  

\bibitem{Enqvist:2001zp} 
  K.~Enqvist and M.~S.~Sloth,
  Nucl.\ Phys.\ B {\bf 626}, 395 (2002)
  [hep-ph/0109214].

\bibitem{Lyth:2001nq} 
  D.~H.~Lyth and D.~Wands,
  Phys.\ Lett.\ B {\bf 524}, 5 (2002)
  [hep-ph/0110002].
 
\bibitem{Moroi:2001ct} 
  T.~Moroi and T.~Takahashi,
  Phys.\ Lett.\ B {\bf 522}, 215 (2001)
  [Erratum-ibid.\ B {\bf 539}, 303 (2002)]
  [hep-ph/0110096].


\bibitem{Moroi:2002rd} 
  T.~Moroi and T.~Takahashi,
  Phys.\ Rev.\ D {\bf 66}, 063501 (2002)
  [hep-ph/0206026].
  
\bibitem{Lyth:2002my} 
  D.~H.~Lyth, C.~Ungarelli and D.~Wands,
  Phys.\ Rev.\ D {\bf 67}, 023503 (2003)
  [astro-ph/0208055].
  
\bibitem{Lyth:2003ip} 
  D.~HLyth and D.~Wands,
  Phys.\ Rev.\ D {\bf 68}, 103516 (2003)
  [astro-ph/0306500].
 
\bibitem{Ferrer:2004nv} 
  F.~Ferrer, S.~Rasanen and J.~Valiviita,
  JCAP {\bf 0410}, 010 (2004)
  [astro-ph/0407300].
  
 
\bibitem{Beltran:2008aa} 
  M.~Beltran,
  Phys.\ Rev.\ D {\bf 78}, 023530 (2008)
  [arXiv:0804.1097 [astro-ph]].

\bibitem{Moroi:2008nn} 
  T.~Moroi and T.~Takahashi,
  Phys.\ Lett.\ B {\bf 671}, 339 (2009)
  [arXiv:0810.0189 [hep-ph]].

\bibitem{Lemoine:2009is} 
  M.~Lemoine, J.~Martin and J.~'i.~Yokoyama,
  Phys.\ Rev.\ D {\bf 80}, 123514 (2009)
  [arXiv:0904.0126 [astro-ph.CO]].
 
 
 
\bibitem{Holder:2009gd} 
  G.~P.~Holder, K.~M.~Nollett and A.~van Engelen,
  Astrophys.\ J.\  {\bf 716}, 907 (2010)
  [arXiv:0907.3919 [astro-ph.CO]].

 
\bibitem{Gordon:2009wx} 
  C.~Gordon and J.~R.~Pritchard,
  Phys.\ Rev.\ D {\bf 80}, 063535 (2009)
  [arXiv:0907.5400 [astro-ph.CO]].
  
\bibitem{Kawasaki:2011ze} 
  M.~Kawasaki, T.~Sekiguchi and T.~Takahashi,
  JCAP {\bf 1110}, 028 (2011)
  [arXiv:1104.5591 [astro-ph.CO]].
 
\bibitem{Grin:2011tf} 
  D.~Grin, O.~Dore and M.~Kamionkowski,
  Phys.\ Rev.\ D {\bf 84}, 123003 (2011)
  [arXiv:1107.5047 [astro-ph.CO]].
 
 
\bibitem{Grin:2013uya} 
  D.~Grin, D.~Hanson, G.~Holder, O.~Dor? and M.~Kamionkowski,
  Phys.\ Rev.\ D {\bf 89}, 023006 (2014)
  [arXiv:1306.4319 [astro-ph.CO]].
  
 
\bibitem{Ade:2013ydc} 
  P.~A.~R.~Ade {\it et al.}  [Planck Collaboration],
  arXiv:1303.5084 [astro-ph.CO].
 
\end{thebibliography}
\end{document}